# Measuring chromatic aberrations in imaging systems using plasmonic nano-particles

Sylvain D. Gennaro, Tyler R. Roschuk, Stefan A. Maier, and Rupert F. Oulton*

Department of Physics, The Blackett Laboratory, Imperial College London, SW7 2AZ, London UK
*Corresponding author: r.oulton@imperial.ac.uk

**Chromatic aberration in optical systems arises from the wavelength dependence of a glass's refractive index. Polychromatic rays incident upon an optical surface are refracted at slightly different angles and in traversing an optical system follow distinct paths creating images displaced according to color. Although arising from dispersion, it manifests as a spatial distortion correctable only with compound lenses with multiple glasses and accumulates in complicated imaging systems. While chromatic aberration is measured with interferometry[1,2] simple methods are attractive for their ease of use and low cost[3,4]. In this letter we retrieve the longitudinal chromatic focal shift of high numerical aperture (NA) microscope objectives from the extinction spectra of metallic nanoparticles[5] within the focal plane. The method is accurate for high NA objectives with apochromatic correction, and enables rapid assessment of the chromatic aberration of any complete microscopy systems, since it is straightforward to implement.**

A straightforward approach for measuring the longitudinal chromatic aberration of an imaging system involves scanning a pin-hole[3] or the confocal image of an optical fibre[4] through a lens's focal plane so that colours in focus record a stronger transmission. The accuracy of these aperture-based measurements requires spatially filtering colors that are out of focus. Therefore, a smaller aperture should increase the sensitivity of these methods as it acts as a point-spread function for the measurement. The pinhole or fiber used must be of a similar scale to the wavelength to accurately assess the chromatic aberration. However, Bethe has shown that the transmission efficiency through a small aperture is fundamentally limited and scales drastically with the inverse fourth power of the aperture's diameter[6] Furthermore, an aperture optimized at one wavelength, has a restricted bandwidth limited by sensitivity for redder wavelengths and point spread function for bluer wavelengths. Applying Babinet's principle can circumvent these limitations: in a chromatic aberration measurement, we can consider replacing the aperture with a resonant scattering object and measuring extinction instead of transmission. For example, metallic nano-particles are particularly attractive dipole scatterers as they support surface plasmon resonances[7] that provide strong scattering cross sections in the visible and near infrared spectral ranges[5]. Indeed, plasmonic electric and magnetic dipole resonances were incorporated within scanning probe techniques to effectively map the amplitude and phase of the electric and magnetic responses in nano-metric optical fields[8,9]. Furthermore, the natural dipole orientation of plasmonic nano-particles can be used to map the electric field vector of an optical field[9,10]. In this letter, we use the far-field extinction spectrum of a plasmonic particle to examine distortions within the focus of an imaging system; in this case longitudinal chromatic aberration. Although Rayleigh scattering by small particles suggests this approach has a similar limitation to Bethe's, it has been shown theoretically that dipole scattering in a focused beam allows near perfect extinction[11–13]. For example, in this work, we achieve up to 95% extinction from a single particle (See Supplementary Figure S1). We utilize this to show a direct correlation between the focal wavelength and the longitudinal chromatic focal shift by analyzing variations in scanned extinction spectra. The method is straightforward, robust, low-cost, broadband (over 800nm+), and 10 – 20 nanometer sensitive, suitable for assessing chromatic aberration in high numerical aperture apochromatic corrected lenses. As the nanoparticles lie on a microscope slide, our method can be retrofitable on any complete microscope system to assess its accumulated chromatic aberrations (Light source, collimating optics and microscope lenses). This is possible because the metallic nanoparticle directly probes the chromatic aberration at the focus and monitors the variation in the spot sizes. Moreover, it also enables the quality control of the chromatic corrections of immersion microscope objectives easily.

Our measurement approach is illustrated schematically in Fig. 1. White light is collimated and expanded to fill the back aperture of a test objective as shown in Fig 1(a), and then focused down onto a single aluminum disk with a diameter in the range from 260 nm up to 520 nm (See Supplementary Figure S1 for the extinction spectra of the five nano-disks presented in this work). The aluminum nano-disks were defined in arrays on a glass slide. A metallic objective re-collimates any transmitted light, which is then projected onto a spectrometer (See Methods). In aligning the optical system, the particle is adjusted on a closed-loop piezoelectric controlled 3-axis translation stage to maximize the overall extinction. From this central position, the particle is scanned along the optical axis over a 1.5 µm range in 20 nm steps for the apochromat objective (60X, NA 0.95) and over a 15 µm range in 100 nm steps for the achromat objective (40X, NA 0.6).

The extinction spectra of the nano-particle were recorded for each scan position along the optical axis for two test objective lenses: an achromat and an apochromat. For clarity, the spectra shown in Fig. 2 are from only a few scan positions. The broken line shows the envelope curve, corresponding to the combined maximum extinction at each wavelength extracted from the complete scan. This curve represents the intrinsic extinction spectrum of the particle with the effect of chromatic lens aberrations eliminated. At each scan position, chromatic aberration distorts the true extinction spectrum since only up to two (achromat) or three (apochromat) colors can be in focus at the same time. As expected, the achromat strongly distorts the extinction spectra (Fig 2a and Supplementary Figure S2) whereas the apochromat gives a far more accurate representation (Fig 2b). It is important to note that the true resonance wavelength and spectral width of the localized plasmonic resonance is easy to mistake when using an achromatic imaging system. Note also, that the tighter focus of the apochromat allows much larger overall extinction, which is expected for dipole scattering in a tightly focused beam[11–13]. Remarkably, the sensitivity of this technique increases for higher NA lenses, which tend to require finer chromatic aberration correction.

The longitudinal chromatic focal shift is retrieved with minimal numerical processing by simply tracking the correspondence of maximum peak extinction at each wavelength with the spectra taken at each scan position along the optical axis. We can build a 2D extinction map of position versus wavelength and normalize this to the maximum extinction at each wavelength, as shown in the insets of Fig. 3. This removes the influence of the intrinsic extinction spectrum of the particle and leaves only the effect of the varying focal position on the extinction. The longitudinal chromatic focal shift corresponds to the maximum of this map, which simply indicates where the focal spot is smallest at each wavelength. Figures 3a and 3b show the broadband (>500nm+) longitudinal chromatic focal shift of the achromat and apochromat, respectively, with a resolution of <20nm, extracted from the 2D map for various disk sizes. The chromatic focal shift shows agreement between the various disks sizes as well as the scale of the expected aberration correction for achromatic and apochromatic lenses. For smaller disks, there is a deviation in the near IR due to the fact that smaller disks have weaker, blue-shifted extinction spectra. Indeed, the maximum extinction depends on the relative strengths of scattering and extinction cross sections and inversely on the spot area[12,13]. As such, small particles generally have reduced sensitivity. These measurements also agree with the manufacturer's data.

The reader should note that the longitudinal chromatic focal shift, presented in Figure 3a and 3b is extremely broadband (>500nm+), limited only by the spectrum of the light source and the sensitivity of the camera. The bandwidth transmitted through an aperture is limited and distorted as a function of wavelength whereas the resonance of an aluminium particle is extremely broad and with a dipole-like response (200nm – 1300nm+).

The choice of metal in constructing the nano-particles is important as it affects the peak as well as the width of the resonance, both of which impact upon the sensitivity. We have measured the longitudinal chromatic aberrations of the two objectives using gold disks with similar diameters. The same chromatic focal shifts were found for long wavelengths, but the inter-band absorption of the gold near 520 nm prevented focal shift retrieval at bluer wavelengths. While silver disks would allow focal shift retrieval across the visible spectrum, aluminum disks also provide the largest extinction in the visible range due to its high plasma frequency[14]. The lower plasma frequencies of both silver and gold disks lead to lower extinction due to higher absorption to scattering cross section ratios (See Supplementary Figure S3 and S4).

Similar to interferometric and aperture-based methods where fine controls of tilt and positioning are important to either minimize phase delay (Interferometry) or maximize transmission, we have also studied the influence of tilt and beam misalignment. This has generally caused minor discrepancies near the spectral extrema of the measured longitudinal chromatic focal shifts. We also compared our result with an optimization procedure, where the chromatic focal shift is retrieved by optimizing the peak extinction of the particle at each wavelength and recording its subsequent position (See Supplementary Figure S5). The results agree well confirming that, for our imaging system, the method is robust against misalignment.

We assess the second objective's impact on the chromatic aberration result by measuring the chromatic focal shift of the apochromat and the achromat with a metallic chromatic aberration free and an achromat microscope objective (See Supplementary Figure S6). As a result, the second objective only slightly influences the aberrations, which is very positive, given the difference in chromatic scale between the apochromatic correction of the objective under test and the achromat correction of the collecting objective. Our method would work equally well in reflection, by placing a beam splitter before the first objective, and monitoring the peak of reflection instead of transmission bypassing any influence on the chromatic correction of the second objective.

The sensitivity of this technique stems from the ratio between the particle's scattering cross section and the area of the focal spot. Remarkably a tighter focus enhances the sensitivity of the technique making it ideal for fine chromatic measurements on high NA microscopy imaging systems incorporating apochromat or even super-apochromat objectives, which are the most difficult to assess such aberrations. Furthermore, since these resonant metallic particles can be smaller than the diffraction limit of light, they admit a negligible point spread function into the measurement across a broad frequency range. This indicates the possibility not only to assess geometric aberrations, but also the interplay of chromatic and geometric aberrations with plasmonic nano-particles in the future.

**Methods**

In our experiment we used an expanded and collimated supercontinuum white light source generated from a non-linear fiber (femtowhite from NKT photonics) pumped by nominally

>150 fs pulses at 820 nm (Chameleon Ultra II) to fill the back aperture of the test objective. (We note that our method would work equally well with any broadband incoherent light source.) The microscope objective then focuses light onto an individual aluminum particle. The second objective re-collimates the light onto a spectrometer and records the transmission spectrum. The spectrometer consists of an F2 prism, a 75 mm cylindrical lens and a CCD Camera with an octave-spanning spectral range from 0.45 $\mu$m to 1 $\mu$m. (The spectral range limitation here comes from the light source and the sensitivity of the camera.) The apparatus is carefully adjusted to minimize the tilt of the nanoparticle relative to the optical axis so as to align one scan axis of the sample holder to the optical axis. This ensured that the longitudinal displacement of the particle is along the optical axis of the microscope objective in question and therefore minimizes additional transverse chromatic aberration during a scan. Experimentally, a longitudinal displacement of 10nm was sufficient to see a clearly visible change in the extinction spectrum, but steps as small as 5nm could be achieved with relatively inexpensive closed-loop piezo control.

**Funding Information**

UK Engineering and Physical Science Research Council (EP/I004343/1), Marie Curie IRG (PIRG08-GA-2010-277080) and the Leverhulme Trust.

**Acknowledgement**

Thanks to Ara Minassian and John Maxwell for useful discussions.

# References

1. Seong, K. & Greivenkamp, J. E. Chromatic aberration measurement for transmission interferometric testing. *Appl. Opt.* **47,** 6508–11 (2008).

2. Amir, W. *et al.* Simultaneous visualization of spatial and chromatic aberrations by two-dimensional Fourier transform spectral interferometry. *Opt. Lett.* **31,** 2927–9 (2006).

3. Kingslake, R. The Measurement of the Aberrations of a Microscope Objective. *J. Opt. Soc. Am* **26,** 251–253 (1936).

4. Juskaitis, R. & Wilson, T. A method for characterizing longitudinal chromatic aberration of microscope objectives using a confocal optical system. *J. Microsc.* **195 (Pt 1),** 17–22 (1999).

5. Novotny, L. & van Hulst, N. Antennas for light. *Nat. Photonics* **5,** 83–90 (2011).

6. Bethe, H. Theory of Diffraction by Small Holes. *Phys. Rev.* **66,** 163–182 (1944).

7. Maier, S. A. *Plasmonics: fundamentals and applications.* (Springer-Verlag Berlin, 2007).

8. Rotenberg, N. & Kuipers, L. Mapping nanoscale light fields. *Nat. Publ. Gr.* **8,** 919–926 (2014).

9. Bauer, T., Orlov, S., Peschel, U., Banzer, P. & Leuchs, G. Nanointerferometric amplitude and phase reconstruction of tightly focused vector beams. *Nat. Photonics* **8,** 23–27 (2013).

10. Lee, K. G. *et al.* Vector field microscopic imaging of light. *Nat. Photonics* **1,** 53–56 (2007).

11. Zumofen, G., Mojarad, N., Sandoghdar, V. & Agio, M. Perfect Reflection of Light by an Oscillating Dipole. *Phys. Rev. Lett.* **101,** 180404 (2008).

12. Mojarad, N. M., Zumofen, G., Sandoghdar, V. & Agio, M. Metal nanoparticles in strongly confined beams: transmission, reflection and absorption. *J. Eur. Opt. Soc. Rapid Publ.* **4,** 09014 (2009).

13. Gennaro, S. D. *et al.* Spectral interferometric microscopy reveals absorption by individual optical nanoantennas from extinction phase. *Nat. Commun.* **5,** 3748 (2014).

14. Hylton, N. P. *et al.* Loss mitigation in plasmonic solar cells: aluminium nanoparticles for broadband photocurrent enhancements in GaAs photodiodes. *Sci. Rep.* **3,** 2874 (2013).

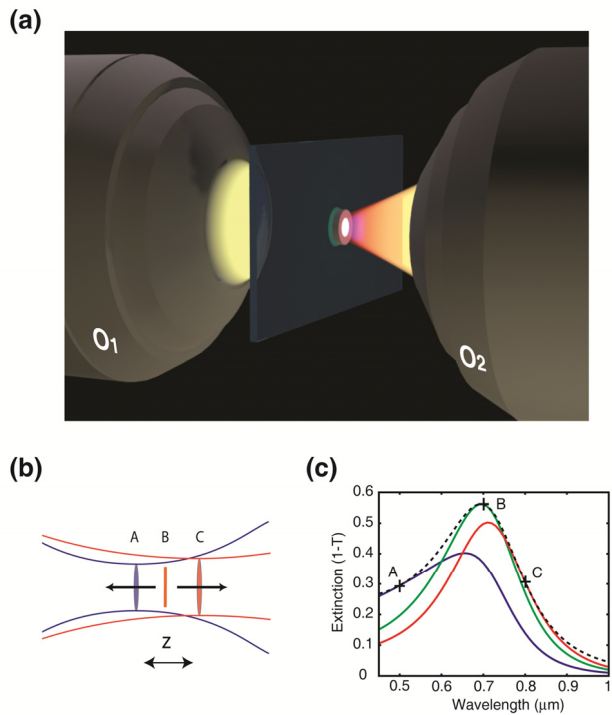

**Fig 1. Measuring longitudinal chromatic aberration with a metallic nano-particle.** (a) Schematic of a chromatic aberration measurement. (b) Illustration of how particle position corresponds to the different extinction spectra in (c) due to chromatic aberrations, (c) Representation of the true extinction spectrum of a single disk for perfect chromatic aberration correction (black curve). Chromatic aberrations distort extinction spectra for various particle positions (red, blue and green) from (b). Wavelengths where a scanned spectrum touches the true extinction spectrum, (A, B and C) are in focus.

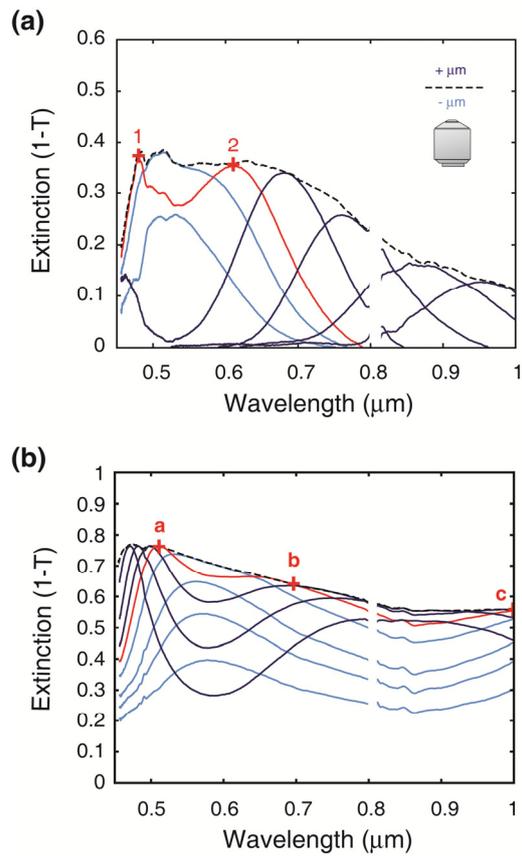

**Fig 2. Comparison of scanned and intrinsic extinction spectra of an aluminum nanoparticle.** Plots are for (a) the achromat and (b) the apochromat objective and a 400 nm diameter aluminum disk. The red curve represents the position of optimal aberration correction. Dark blue (cyan) curves are when the aluminum disk is moved further (closer) to the objective from this optimal position (See Methods). The black dotted line represents the maximum extinction (at each wavelength) taken from all the scans. Labels 1 and 2 (respectively the letters a,b,c) indicate points in focus for the optimal extinction curve (See Fig 3).

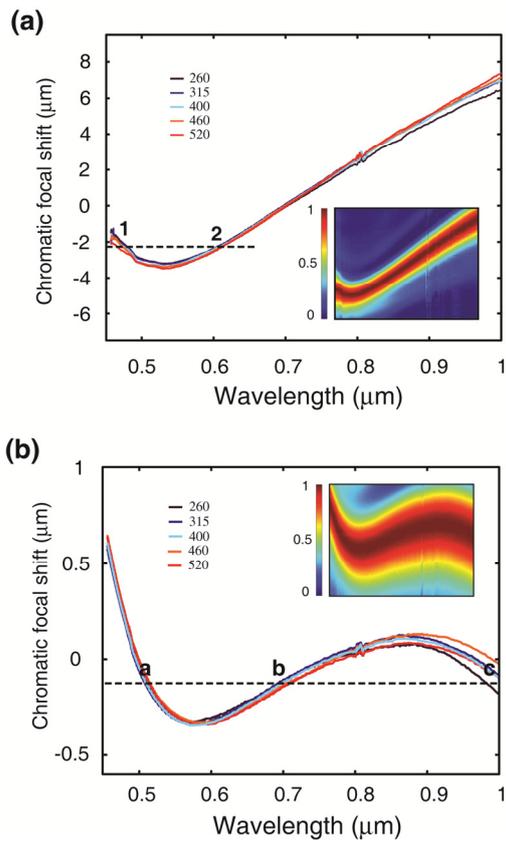

**Fig 3. Measured longitudinal chromatic focal shifts.** Focal shifts are plotted for (a) the achromat and (b) the apochromat objective scanned using aluminum disks of varying size. The insets show the normalized extinction maps of position (20nm resolution) versus wavelength for a 400 nm diameter aluminum disk. Labels 1 and 2 (respectively, letters a, b and c) indicate the point in focus for the optimal extinction curve (see Fig 2).